\documentclass[aps,prd,amsmath,floats,floatfix,twocolumn,nofootinbib]{revtex4}
\usepackage{graphicx}
\usepackage{bm}

\renewcommand{\Re}{\mathrm{Re}}

\begin{document}
\vspace{-2.5cm} 

\title{Solving Einstein's Equations With Dual Coordinate Frames}

\author{Mark A. Scheel${}^1$, Harald P. Pfeiffer${}^1$, Lee Lindblom${}^1$, 
 Lawrence E. Kidder${}^2$,\\ Oliver Rinne${}^1$, and Saul A. Teukolsky${}^2$}

\affiliation{${}^1$Theoretical Astrophysics 130-33, California Institute of
Technology, Pasadena, CA 91125}

\affiliation{${}^2$Center for Radiophysics and Space Research, 
Cornell University, Ithaca, New York, 14853}

\date{\today}

\begin{abstract}
A method is introduced for solving Einstein's equations using two
distinct coordinate systems.  The coordinate basis vectors associated
with one system are used to project out components of the metric and
other fields, in analogy with the way fields are projected onto an
orthonormal tetrad basis.  These field components are then determined
as functions of a second independent coordinate system.  
The transformation to the second coordinate system can be thought of
as a mapping from the original ``inertial''
coordinate system to the computational
domain. This
dual-coordinate method is used to perform stable numerical evolutions
of a black-hole spacetime using the generalized harmonic form of Einstein's
equations in coordinates that rotate with respect to the
inertial frame at infinity; such evolutions are found to be
generically unstable using a single rotating coordinate frame.  The
dual-coordinate method is also used here to evolve binary black-hole
spacetimes for several orbits.  The great flexibility of this method
allows comoving coordinates to be adjusted with a feedback control
system that keeps the excision boundaries of the holes within their
respective apparent horizons.
\end{abstract}


\maketitle

\section{Introduction}
\label{s:Introduction}

This paper introduces a new method of solving Einstein's equations
using two distinct coordinate systems.  Tensors like the metric are
represented by their components in the coordinate basis of the first
coordinate system.  Einstein's equations are then used to determine
these tensor components as functions of the second coordinate
system. The mapping between the
first and second coordinate systems is quite arbitrary and can be
chosen dynamically.  This freedom allows us to adapt the second
coordinate system continuously, for example to track the motion of the
individual black holes in a binary.
The first coordinate system plays
much the same role as the orthonormal tetrad used in some formulations
of the Einstein equations~\cite{estabrook_etal97, FriedrichNagy1999,
Buchman2003}.  The transformation to the
second coordinate system can be thought of as a
mapping between the original ``inertial'' coordinates and the
computational domain, and plays much the same role as the ``grid
velocity'' sometimes used in numerical
hydrodynamics~\cite{Wilson1979}.  The bulk of this paper gives a
careful description of this new method as it applies to the
generalized harmonic form of the Einstein
equations~\cite{Lindblom2006}, plus a set of numerical tests that
demonstrate its usefulness.

The flexibility to choose a coordinate frame adapted to a particular
physical problem is often used to simplify solutions of the Einstein
equations.  For example, solutions having some symmetry ({\it e.g.,}
time-independence) are much simpler when expressed in coordinates that
respect that symmetry.  It has long been expected that
binary black-hole spacetimes would most naturally be represented in a
coordinate frame that corotates with the orbit of the holes; this
frame should be advantageous for numerical simulations since it would
make the fields nearly time-independent during the inspiral.
Unfortunately our attempts to model binary black-hole spacetimes in
the conventional single-coordinate framework have all failed.  These
failures reveal several serious and interesting problems with the
single-frame approach, and motivated us to develop the new methods
described in this paper.  Before we turn to our discussion
of the new dual-frame approach, however, we believe it is useful to
describe those problems with the standard single-frame
methods.

The groundbreaking binary black-hole evolutions of
Pretorius~\cite{Pretorius2005a} use a single coordinate frame that
asymptotically approaches the inertial frame at infinity. The
singular interior regions of the holes are excised from the
computational domain.  In this approach the black holes move across
the coordinate grid and the solution is time dependent on the
orbital timescale.  We have attempted to implement a similar scheme
using our spectral evolution code, but encountered two significant
problems: First, the black-hole ``excision'' boundaries must be
changed from time to time to track the motions of the black-hole 
horizons.  This requires finding a way of
adding (and deleting) grid points to (and from) the computational
domain in a way that keeps the distance between the horizon and the
excision boundary more or less fixed, and then---the difficult part---to
find sufficiently smooth and stable ways to fill ({\it e.g.,} by
extrapolation) the new grid points with appropriate values for all the
dynamical fields.  This problem has been solved successfully for
lower-order finite-difference numerical methods~\cite{bbhprl98a,
Brandt2000, Alcubierre2001, Shoemaker2003, Pretorius2005a}.  The
difficulty for the spectral-method implementation seems to be the
inability to produce sufficiently smooth and well-behaved
extrapolations of the needed dynamical field quantities to points
lying beyond the current computational domain.  Our attempts
to develop such techniques have not been successful.

There is also a second fundamental problem with the conventional
moving-excision method: although the black-hole horizons move through
the computational grid, the excision boundaries are fixed except when
they are moved to update the grid. To understand why this can cause
trouble, consider first an excision surface located just inside the
horizon, but {\em moving continuously} along with the horizon.  In
this case, the excision boundary is a spacelike surface that lies in
the future of the computational domain, so (for a causal hyperbolic
representation of the Einstein equations) boundary conditions are not
needed there, and the spacetime region inside can be excised.  Now
consider the conventional moving-excision method, where the boundaries
are {\em fixed} in the coordinate grid during a time step, and
therefore move relative to the horizons.  At the trailing edge of the
horizon, the excision surface is moving out of the black hole
superluminally. Locally this surface is spacelike and lies in the
future of the computational domain, so boundary conditions are not
needed there.  But at the leading edge, the excision surface is
falling deeper into the black hole; if it falls quickly enough, some
part of this surface may become timelike, so boundary conditions
will be required at these points.  If appropriate
boundary conditions are not supplied, the evolution
problem becomes ill-posed and the black-hole excision paradigm breaks
down.  We monitor the characteristic speeds at excision
boundaries in our code, and have verified that this breakdown
does occur in our simulations
when a black hole moves too quickly through the coordinate grid.
This problem can be ameliorated by moving the excision
boundaries deeper into the black-hole interiors (where with an
appropriate choice of gauge, light cones are ``tipped'' further toward
the singularity); we presume this is how the successful
finite-difference codes control this problem.  But we were not
successful in curing this problem in our spectral evolution code.
Another successful approach to moving black holes through the
computational grid is the moving-puncture method, {\it cf.}
Refs.~\cite{Campanelli2006a, Baker2006a, Campanelli2006b, Baker2006b,
Campanelli2006c, vanMeter2006}, which can be thought of as shrinking
the internal excision boundaries to isolated points 
that lie between the grid points (and so are simply ignored).  
Unfortunately the
dynamical fields are not smooth at the puncture points, so the moving
puncture approach may not be extendable to spectral methods.

\begin{figure} 
\begin{center}
\hbox{\hspace{.50cm}
\includegraphics[width=3.0in]{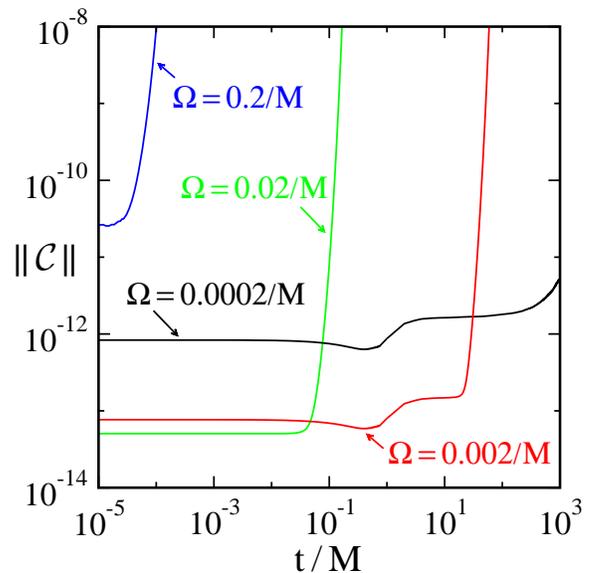}}
\end{center}
\vspace{-.95cm}
\caption{Constraint violations in evolutions of a Schwarzschild black
hole using the generalized harmonic system and
coordinates that rotate with angular velocity $\Omega$.
\label{f:fig1}}
\vspace{-0.25cm}
\end{figure}

Since we were unsuccessful in moving black holes through our spectral
computational domain using the conventional method, we were led to
pursue an approach using comoving coordinates where the black holes
remain at rest.  Unfortunately, the generalized harmonic evolution
system~\cite{Lindblom2006} exhibits severe instabilities 
(in the sense that the constraints grow without bound) when initial
data are evolved using coordinates that rotate with respect to the
inertial frame at infinity.\footnote{This instability appears to be a
feature of the generalized harmonic system.  No
instability was seen in earlier evolutions
of black holes in rotating frames using other
forms of the Einstein system~\cite{Gomez1998}, and
we do not find such an instability using our code with the KST
evolution system~\cite{Kidder2001}.  In the GH system the full
4-metric, including the lapse and shift, are evolved as
dynamical fields, while in the KST system only the 3-metric
is evolved. Since only the shift grows near spatial infinity in a rotating 
frame, the non-dynamical treatment of the shift in the KST system may
account for its rotating frame stability. }  
These
instabilities occur even for simple time-independent cases like
Minkowski or Schwarzschild spacetime.  For example, Fig.~\ref{f:fig1}
shows the constraint violations for evolutions of Schwarzschild in a
frame rotating with angular velocity $\Omega$.  These evolutions were
performed on a computational domain that is a little larger than the
size needed for binary black-hole evolutions, with outer radius
$R_{\mathrm{max}}=1000M$; see Sec.~\ref{s:RotatingFrame} for details.
Binary black-hole evolutions must be performed stably and accurately
for times of order $1000M$ to model all of the interesting inspiral
and merger dynamics.  However, the rotating frame evolution shown in
Fig.~\ref{f:fig1} with $\Omega=0.2/M$ (comparable to the maximum
angular velocity achieved by a black-hole binary just before merger)
is unstable in the unacceptably short time of about $10^{-4}M$.  We do
not fully understand the instability in evolving asymptotically flat
spacetimes in rotating coordinates using the generalized harmonic
evolution system.  We have reduced the severity of this problem with
methods discussed in the Appendix, but we were not able to solve
completely the problem of evolving with a single corotating coordinate
frame.

Another fundamental limitation on using rotating coordinates in
asymptotically-flat spacetimes is a numerical resolution issue: some
components of the four-metric grow asymptotically like
$\rho^2\Omega^2$ for large values of the cylindrical radial coordinate
$\rho$ (to leading order), while other components approach $M/r$, for
large values of the spherical radial coordinate $r$.  On computational
domains that extend to $r\approx 1000M$ and angular velocities
$\Omega\approx 0.2/M$ (appropriate for binary black-hole spacetimes),
these values differ by almost eight orders of magnitude, so that the
dynamical range needed to resolve the various field components with
sufficient accuracy is difficult to achieve using double precision
numerical methods.  This difficulty goes away if one uses the inertial
frame components of the various fields~\cite{Pretorius2005a}.

The dual-frame approach proposed here corrects or circumvents all of
the problems associated with the more traditional approaches that we
know about.  The remainder of this paper is organized as follows.  The
dual-coordinate frame method is presented more completely in
Sec.~\ref{s:Moving Frames}.  This new method is applied in
Sec.~\ref{s:RotatingFrame} to the case where the two coordinate frames
rotate uniformly with respect to each other.  We test this new
rotating-coordinate method by evolving Schwarzschild initial data in
coordinates that rotate with respect to infinity, finding that the
evolutions of Fig.~\ref{f:fig1} become stable and convergent.  A
number of technical numerical issues associated with the
dual-coordinate method ({\it e.g.,} how to construct the appropriate
spectral filters) are also discussed in Sec.~\ref{s:RotatingFrame}.
Section~\ref{s:BBHtests} further tests this new method by evolving a
binary black-hole spacetime in uniformly rotating coordinates.  This
test fails when the fixed angular velocity of the rotating frame no
longer tracks the motion of the black holes to sufficient accuracy.
In Sec.~\ref{s:HorizonTracking} we present a more sophisticated
application of the dual-coordinate method by constructing a flexible
coordinate map whose parameters are adjusted through a feedback
control system that keeps the black-hole horizons centered on the
excsion surfaces.  Using these new horizon-tracking coordinates, we are
able to evolve a binary black-hole spacetime in a stable and
convergent manner for about 4.6 orbits.  Our results are
summarized, and a discussion of possible directions for further
development are given in Sec.~\ref{s:Discussion}.  Finally, we review
in the Appendix our attempts to make rotating
single-coordinate frame evolutions stable for the generalized harmonic
evolution system.

\section{Dual Coordinate Frames}
\label{s:Moving Frames}

Consider a first-order representation of the Einstein evolution system,
such as our formulation of the generalized harmonic 
system~\cite{Lindblom2006}.  The evolution equations for the dynamical
fields $u^{\bar \alpha}$ for such systems can be represented
abstractly as
\begin{eqnarray}
\partial_{\kern 0.1em \bar t} u^{\bar \alpha} 
+ A^{\bar k\,\bar \alpha}{}_{\bar
\beta} \partial_{\kern 0.1em \bar k} u^{\bar \beta} = F^{\bar\alpha},
\label{e:evolsysdef}
\end{eqnarray}
where $A^{\bar k\,\bar \alpha}{}_{\bar \beta}$ and $F^{\bar \alpha}$
may depend on $u^{\bar \alpha}$ but not its derivatives.  We use Greek
indices $\bar\alpha$, $\bar \beta$, ... to label the various dynamical
fields, and Latin indices $\bar \imath$, $\bar \jmath$, $\bar k$, ... to label
the spatial coordinates.  The bars on the various indices in
Eq.~(\ref{e:evolsysdef}) indicate that this system evolves the
coordinate components of the collection of dynamical fields, $u^{\bar
\alpha}$, as functions of coordinates $x^{\bar a}=(\bar t,x^{\bar
\imath})$.  In the generalized harmonic system, for example, $u^{\bar
\alpha} = \{\psi_{\bar a\bar b}, \Pi_{\bar a \bar b}, \Phi_{\bar k
\bar a\bar b}\}$ where $\psi_{\bar a\bar b}$ is the four-metric,
$\Pi_{\bar a\bar b} = - t^{\bar c}\partial_{\kern 0.1 em \bar c}
\psi_{\bar a \bar b}$, $\Phi_{\bar k\bar a\bar b}= \partial_{\kern 0.1
em \bar k}\psi_{\bar a\bar b}$, and $t^{\bar c}$ is the unit timelike
normal to the $\bar t =$constant hypersurfaces.  We use Latin indices
from the beginning of the alphabet, $\bar a$, $\bar b$, $\bar c$, ...
to label spacetime coordinates.

Einstein's equations are covariant, so it is straightforward to
transform any representation of those equations from one coordinate
frame to another.  The standard method of transformation changes both
the coordinates, $x^{\bar a}\rightarrow x^a$, and the components of
the dynamical fields, $u^{\bar \alpha}\rightarrow u^\alpha$, using the
appropriate transformation rules.  For the generalized harmonic
system, for example, the field components transform according to the
rules:
\begin{eqnarray}
\psi_{ab} &=& \frac{\partial x^{\bar a}}{\partial x^a}
\frac{\partial x^{\bar b}}{\partial x^b}\psi_{\bar a\bar b},\\
\label{e:pitrafo}
\Pi_{ab} &=& \frac{\partial x^{\bar a}}{\partial x^a}
\frac{\partial x^{\bar b}}{\partial x^b}\Pi_{\bar a\bar b}
-2 \frac{\partial^{\kern 0.1em 2} x^{\bar a}}{\partial x^a \partial x^c}
\frac{\partial x^{\bar b}}{\partial x^b}
\frac{\partial x^{c}}{\partial x^{\bar c}}\psi_{\bar a\bar b}t^{\bar c},\\
\label{e:phitrafo}
\Phi_{kab} &=& \frac{\partial x^{\bar k}}{\partial x^k}
\frac{\partial x^{\bar a}}{\partial x^a}
\frac{\partial x^{\bar b}}{\partial x^b}\Phi_{\bar k\bar a\bar b}
+2\frac{\partial^{\kern 0.1 em 2} x^{\bar a}}{\partial x^k \partial x^a}
\frac{\partial x^{\bar b}}{\partial x^b}\psi_{\bar a\bar b}.
\end{eqnarray}

Here we propose a new way of solving the Einstein equations that
solves both the rotating frame instability problem and the
moving-excision-boundary problems discussed in
Sec.~\ref{s:Introduction}.  We introduce a second coordinate system
$x^a$ that is (in principle) completely independent of the first
$x^{\bar a}$.  We think of the first coordinate system $x^{\bar a}$ as
inertial coordinates that do not rotate with respect to the asymptotic
inertial frame at spatial infinity.  We use these inertial coordinate
bases, $\partial_{\bar a}$ and $dx^{\bar a}$, to construct the
components of the various dynamical fields: $u^{\bar\alpha}$.  These
inertial-frame components are well behaved near spatial infinity, and
the numerical dynamical range needed to represent them is
significantly reduced.  We think of the second set of coordinates,
$x^a$, as comoving coordinates chosen to minimize the time dependence
of the dynamical fields in some way.  We solve the evolution system
for the inertial-frame dynamical-field components $u^{\bar\alpha}$ as
functions of the comoving coordinates $x^a$.  For simplicity we
consider only the case where the two coordinate systems have the same
time slicing: $t=\bar t$.  In this case the system of evolution
equations for $u^{\bar\alpha}$ in terms of $x^a$ is just
Eq.~(\ref{e:evolsysdef}) with the straightforward change of
independent variables:
\begin{eqnarray}
\partial_{t} u^{\bar \alpha} + \biggl[
\frac{\partial x^i}{\partial \bar t}\delta^{\bar\alpha}{}_{\bar\beta}
+\frac{\partial x^i}{\partial x^{\bar k}}
A^{\bar k\,\bar \alpha}{}_{\bar
\beta} \biggr]
\partial_{i} u^{\bar \beta} = F^{\bar\alpha}.
\label{e:newevolsysdef}
\end{eqnarray}
Here $\partial x^i/\partial \bar t$ and $\partial x^i/\partial x^{\bar
k}$ are to be determined as functions of $t$ and $x^i$ from the
transformation that relates the two coordinate systems: $x^i =
x^i(\bar t, x^{\bar k})$.  

The dual-coordinate evolution system,
Eq.~(\ref{e:newevolsysdef}), has many properties in common with the
original, Eq.~(\ref{e:evolsysdef}).  Its solutions are the same
(assuming the coordinate transformation is sufficiently smooth), just
expressed in terms of the new coordinates $x^a$:
$u^{\bar\alpha}(x^a)=u^{\bar\alpha}[x^{\bar b}(x^a)]$. In addition the
characteristic fields are exactly the same for the two systems, and
the characteristic speeds differ only by 
$v = \bar{v} + n_i \partial x^i/\partial \bar t$, 
where $\bar{v}$ represents any of the characteristic speeds of
Eq.~(\ref{e:evolsysdef}) and $v$ is the corresponding characteristic
speed of Eq.~(\ref{e:newevolsysdef}).
Here $n_i$ is the appropriate normal one-form used to define
the characteristic speeds ({\it e.g.,} the outward-directed normal to
the boundary surface).  Thus boundary conditions for the two systems
are transformed in the obvious way.  Since the characteristic speeds
may be different, however, it is possible that the list of
characteristic fields needing boundary conditions at a particular
point may change.

\section{Rotating Coordinate Frames}
\label{s:RotatingFrame}

In this section we present a simple test of the dual-coordinate-frame
idea described in Sec.~\ref{s:Moving Frames}.  We evolve
asymptotically flat spacetimes, like the Minkowski and Schwarzschild
geometries, using two frames: an inertial coordinate frame that is
asymptotically Cartesian at infinity, and a comoving coordinate frame
that rotates uniformly with respect to the inertial frame.  

Before we discuss those
tests, however, it is appropriate to describe the numerical methods
that we use.  These tests are done with the generalized harmonic
evolution system as described in Ref.~\cite{Lindblom2006} using
spectral numerical methods as described, for example,
in Refs.~\cite{Kidder2005, Boyle2006}.  
The components of the various fields in these tests are
expanded in terms of scalar spherical harmonics of the angular coordinates
(with $\ell\leq 11$) and Chebyshev polynomials of the radial
coordinate $\log r$ through order $N_r-1$ (with $N_r=15$ for the tests
in Fig.~\ref{f:fig1}).  The computational domain used in these tests
(as well as those shown in Fig.~\ref{f:fig1}) consists of a set of
eight nested spherical shells with boundaries located at the radii
1.8, 8, 35, 70, 140, 229, 374, 612, and $1000M$ respectively.  We
measure constraint violations in these tests with a quantity $||{\cal
C}||$ defined as the ratio of the $L^2$ norm of all the constraint
fields of the generalized harmonic system, divided by the $L^2$ norm
of the spatial gradients of the various dynamical fields $\partial_k
u^{\bar\alpha}$ (see Eq.~[71] of Ref.~\cite{Lindblom2006}).  This
quantity $||{\cal C}||$ vanishes whenever the constraints are
satisfied, and the normalization is chosen so that $||{\cal C}||$
becomes of order
unity when constraint violations dominate the solution.

In 3+1 formulations of general relativity, the gauge freedom in the
theory is usually parameterized by a lapse function and a shift vector
that are freely specifiable.  In contrast, the gauge freedom in
generalized harmonic formulations of Einstein's equations is
represented by four freely-specifiable gauge source functions $H_a$
(see Ref.~\cite{Lindblom2006}) which determine the evolution of the
lapse and shift. Once $H_a$ has been chosen, the harmonic
constraint equation
\begin{equation}
  \label{e:ghconstr}
  0 = \mathcal{C}_a \equiv \Gamma_a + H_a
\end{equation}
must be satisfied. Here $\Gamma_a \equiv \psi^{bc}\Gamma_{abc}$ is a
trace of the Christoffel symbols.  For single-coordinate-frame
numerical tests that aim to reproduce a known analytic
time-independent solution, we typically choose $H_a$ so that the
constraint, Eq.~(\ref{e:ghconstr}), is satisfied initially, and we fix
this $H_a$ for all time; this was done for example in the evolution shown in
Fig.~\ref{f:fig1}.

For dual-frame evolutions we must be more careful with the gauge
source functions, particularly because $H_a$ does not transform like a
tensor.  Because we have had considerable success (in nonrotating
frames) choosing $H_a$ to be time-independent, we take a similar
approach here. We first define a new quantity $\tilde{H}_a$ that has
the following two properties: 1) $\tilde{H}_a$ transforms like a
tensor, and 2) in inertial coordinates $\tilde{H}_{\bar a} = H_{\bar
a}$.  As in the single-frame case, we choose $H_a$ so that the
constraint Eq.~(\ref{e:ghconstr}) is satisfied initially, but now we
demand that $\tilde{H}_a$ is constant in the moving frame, {\it i.e.,}
that $\partial_t \tilde{H}_a = 0$.

\begin{figure} 
\begin{center}
\hbox{\hspace{.50cm} \includegraphics[width=3.0in]{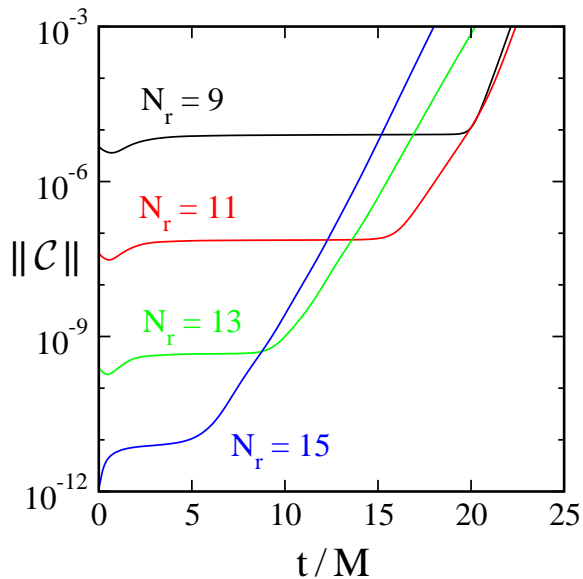} }
\end{center}
\vspace{-0.95cm}
\caption{Constraint violations in dual-coordinate-frame evolutions
of Schwarzschild, with comoving coordinates that rotate uniformly
with angular velocity $\Omega=0.2/M$.  Angular filtering (see text) is 
performed on the inertial frame components.
\label{f:fig2}}
\vspace{-0.25cm}
\end{figure}

Our first test of the dual-coordinate-frame idea consists of evolving
Schwarzschild initial data with uniformly rotating coordinates.  In
particular we repeat the most unstable evolution shown in
Fig.~\ref{f:fig1} as a dual-coordinate-frame evolution.  The inertial
coordinates $(\bar t, \bar x, \bar y, \bar z)$ for this test are the
standard asymptotically Cartesian coordinates associated with the
Kerr-Schild representation of the Schwarzschild geometry, while the
``comoving'' coordinates $(t,x,y,z)$ rotate uniformly with respect to
these inertial coordinates:
\begin{eqnarray}
\label{e:rotfirst}
t&=&\bar t,\\
x&=& \bar x \cos(\Omega \bar t\kern 0.1em) 
+ \bar y \sin (\Omega \bar t\kern 0.1em),\\
y&=& -\bar x \sin(\Omega \bar t\kern 0.1em) 
+ \bar y \cos (\Omega \bar t\kern 0.1em),\\
\label{e:rotlast}
z&=& \bar z.
\end{eqnarray}
(The angular velocity, $\Omega$, of these co-moving
coordinates can be chosen arbitrarily.) 
We solve the dual-coordinate form of the evolution equations,
Eq.~(\ref{e:newevolsysdef}), for these Schwarzschild initial data with
$\Omega=0.2/M$.  This value of $\Omega$ is chosen
because it corresponds roughly to the orbital angular
velocity of a binary black-hole system at the time of merger.
Figure~\ref{f:fig2} shows the constraint violations
for these evolutions for several values of the radial resolution
parameter $N_r$.  These dual-coordinate-frame evolutions persist for
about $20M$, five orders of magnitude longer than the analogous single
coordinate evolution shown in Fig.~\ref{f:fig1}.  Unfortunately these
evolutions are still unstable on a timescale that is several orders of
magnitude shorter than needed, and this instability shows some
signs of non-convergence.

\begin{figure} 
\begin{center}
\vspace{0.2cm}
\hbox{\hspace{0.25cm}
\includegraphics[width=3.0in]{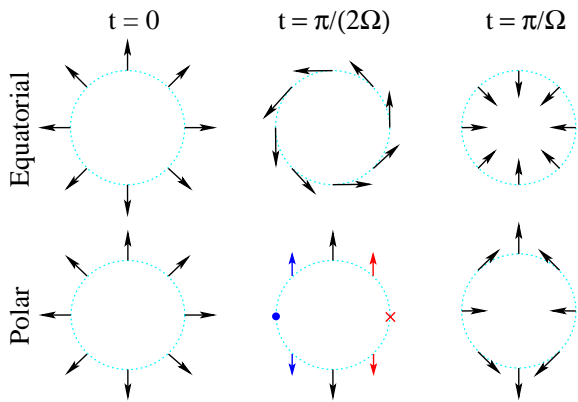}
}
\end{center}
\vspace{-0.95cm}
\caption{Time-dependence of the inertial frame components of two
cross sections (equatorial and polar) of a 
vector field expressed as functions of rotating coordinates.
At $t=0$ the vector field is radial, but
the $\partial_{\kern 0.1em \bar x}$ and
$\partial_{\kern 0.1em \bar y}$ basis vectors rotate
with angular velocity $\Omega$.  The vectors on the right side of the
polar cross section figure at $t=\pi/(2\Omega)$ are pointed into the plane 
of the figure, while those on the left point out.
\label{f:fig3}}
\vspace{-0.25cm}
\end{figure}

The dual-coordinate-frame representation of tensor fields has some
unpleasant features that are ultimately responsible for the
instability seen in Fig.~\ref{f:fig2}.  If the transformation between
the two coordinate frames $x^i = x^i(\bar t,x^{\bar \jmath})$ is
time-dependent (as is the case for uniform rotation), then clearly the
transformation relating the tensor bases, {\it e.g.,} $\partial_{\kern
0.1em \bar \imath}= \partial x^j/\partial x^{\bar\imath}\,\partial_{j}$, will
also be time-dependent.  So even if the comoving components of a
tensor $\psi_{ab}(x^j)$ are {\it time-independent}, as is the case for
the Schwarzschild example above, then the inertial frame components
$\psi_{\bar a\bar b}(t,x^j)$ that we are evolving will change with
time.  Consider for example a purely radial vector field 
$v^i(x^j)$
that is
time-independent when represented in a single coordinate frame that
rotates with angular velocity $\Omega$.  Figure~\ref{f:fig3}
illustrates the time-dependence of the inertial frame components
$v^{\bar \imath}(x^j)$ of this vector field.  Two cross sections of this
field (one equatorial and one polar) are shown at time $t=0$, when the
rotation map is just the identity, and at times $t=\pi/(2\Omega)$ and
$t=\pi/\Omega$, when each inertial frame basis vector
$\partial_{\kern 0.15em \bar \imath}$ is rotated by $\pi/2$ and $\pi$
about the $z$ axis respectively.

\begin{figure}
\vspace{0.2cm} 
\begin{center}
\hbox{\hspace{.50cm}
\includegraphics[width=3.0in]{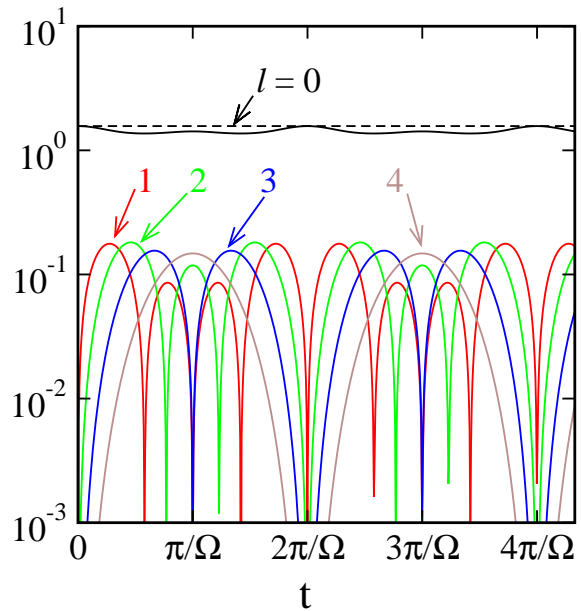}
}
\end{center}
\vspace{-0.95cm}
\caption{Tensor-spherical-harmonic components of the Schwarzschild
spatial three-metric expressed in rotating coordinates.  Dashed curves show the
components expressed in rotating-frame tensor spherical harmonics
(only the $\ell=0$ component is non-zero); solid curves show the
inertial-frame tensor-spherical-harmonic components.
\label{f:fig4}}
\vspace{-0.25cm}
\end{figure}

In addition to making tensor field components time-dependent, the
dual-coordinate-frame representation can also mix the tensor-spherical-harmonic
components of tensors.  In Fig.~\ref{f:fig3} for
example, the radial vector field is pure $\ell=0$ when represented in
rotating-frame vector spherical harmonics, but it becomes a time-dependent
mixture of $\ell=0$, 1, and 2 when represented in
inertial-frame vector spherical harmonics.  Figure~\ref{f:fig4}
illustrates the time dependence of the tensor-spherical-harmonic
components of the Schwarzschild spatial three-metric expressed in rotating 
coordinates. The dashed curves are the
rotating-frame tensor-spherical-harmonic components, while the solid
curves are the inertial-frame tensor-spherical-harmonic components.
In the rotating frame the Schwarzschild spatial three-metric is pure 
$\ell=0$, while
the inertial frame components are a time-dependent mixture of
$\ell=0$, 1, 2, 3, and 4.  The periodicity of the tensor-spherical-harmonic 
components in Fig.~\ref{f:fig4} is determined by the rotation
period of the coordinates $2\pi/\Omega$.

This mixing of the inertial-frame tensor-spherical-harmonic components
is the cause of the instability seen in Fig.~\ref{f:fig2}.  When using
spectral methods in a spherical computational domain, we expand the
Cartesian components of tensors in {\it scalar} spherical harmonics, and we
evolve these Cartesian components.  In this case we find it necessary
periodically to set to zero the highest-order
{\it tensor}-spherical-harmonic coefficients of all tensors in order to avoid
numerical instabilities~\cite{Holst2004,Kidder2005}.  We call this
operation ``filtering'', since it is similar to the filtering operations
commonly used in spectral methods to avoid instabilities caused
by aliasing.  Filtering is necessary because spatial differentiation 
couples different values of the $\ell$ and $m$ indices
in the {\it scalar}-spherical-harmonic expansions of the
Cartesian components of
tensors.  Consequently any truncated series expansion will result in
incorrect evolution equations for the highest angular modes, and these
errors often lead to non-convergent instabilities.  
However, it is possible to truncate the {\it tensor}-spherical-harmonic
expansion of a tensor at a finite value of $\ell$ in
a self-consistent way, because  the spatial gradient of a
tensor spherical harmonic is also a tensor spherical harmonic with the
same spherical harmonic index, but one higher tensor rank. Thus we perform
filtering by transforming each tensor from a Cartesian-component basis
to a tensor-spherical-harmonic basis, zeroing the tensor-spherical-harmonic
coefficients for values of $\ell$ larger than those kept in our expansion, 
and transforming back to our Cartesian-component basis.
This filter cures the angular instability
problems associated with the evolutions of single-coordinate frame
spherical-harmonic representations of
tensors~\cite{Holst2004,Kidder2005}.

\begin{figure} 
\begin{center}
\vspace{0.2cm}
\hbox{\hspace{.50cm}
\includegraphics[width=3.0in]{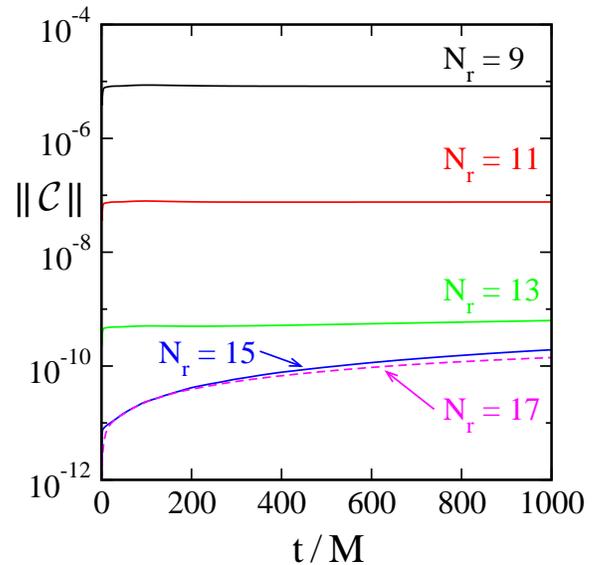}
}
\end{center}
\vspace{-0.95cm}
\caption{Constraint violations in dual-coordinate-frame evolutions
of Schwarzschild, with comoving coordinates that rotate uniformly
with angular velocity $\Omega=0.2/M$.  These evolutions use the new
rotating-frame tensor-spherical-harmonic filtering algorithm.
The highest resolution case (dashed curve) uses more angular
basis functions ($\ell\leq 13$ instead 
of $\ell\leq 11$) as a more stringent test of convergence.
\label{f:fig5}}
\vspace{-0.25cm}
\end{figure}

This filtering algorithm must be modified for dual-coordinate-frame
methods because there is now a more complicated relationship between
the coordinates and the basis vectors used to represent tensors.
Figures~\ref{f:fig3} and \ref{f:fig4} illustrate that inertial-frame
components of a tensor expressed as functions of rotating-frame
coordinates have some additional time-dependent mixing among the
various tensor-spherical-harmonic components.  Therefore, filtering
the inertial-frame components using the same algorithm used for
single-coordinate frame evolutions is the wrong thing to do: it does
not preserve all of the information needed to determine a number of
the highest-index coefficients in this case.  This straightforward
(and incorrect) implementation of spherical-harmonic filtering is the
method used for the test shown in Fig.~\ref{f:fig2}, with the result
being unstable (in the sense that the constraints grow without bound)
and probably non-convergent.  The cure for this problem is also clear:
transform spatial tensors to a {\it rotating-frame}
tensor-spherical-harmonic basis before filtering, then transform back
to the inertial-frame basis afterwards.  (Spacetime tensors, such as
the four-metric $\psi_{\bar a\bar b}$, are first split into their
spatial-tensor parts, {\it e.g.,} $\psi_{\bar t\bar t}$, $\psi_{\bar
t\bar \imath}$, and $\psi_{\bar \imath\bar \jmath}$, and these parts
are then filtered in this way.)  Results using this new filtering
algorithm are shown in Fig.~\ref{f:fig5} for the same initial data and
dual-coordinate frames used in Fig.~\ref{f:fig2}.  These evolutions
now appear to be stable and convergent on the needed timescale, except
for mild (sublinear) power-law growth seen here only in the highest
resolution cases.  We suspect that this power-law growth may be due to
accumulated
roundoff error, but we have not investigated this in detail because
the growth is insignificant on the timescale needed for
multiple-orbit binary black hole evolutions such as those described in
Sec.~\ref{s:HorizonTracking}.
The new filtering method described here  
is needed to perform stable
evolutions of any rotating spacetime (even ``linear'' problems like
evolutions of Minkowski spacetime) using the dual frame method.
Additional filtering may also be needed in some circumstances to
control other problems (spectral aliasing errors due to 
non-linearities for example),
but no additional filtering was needed or used for any of the
numerical evolutions presented in this paper.
%

\section{Binary Black Hole Tests}
\label{s:BBHtests}

The primary motivation for developing the dual-coordinate method
describe in Sec.~\ref{s:Moving Frames} was to allow us to perform
binary black-hole evolutions using coordinates that move with the
holes.  In this section we describe the first test of this new method
for binary black-hole evolutions.  We describe the binary black-hole
initial data that we use, give a very brief description of the
binary-black-hole specific features of our numerical methods, and then
describe the results of this first test.

The binary black-hole evolutions described here begin with initial
data prepared with the methods described in Ref.~\cite{Cook2004}.
These data represent two equal-mass corotating black holes in a
quasi-stationary circular orbit.  The data are obtained by solving the
extended conformal  thin-sandwich form of the initial value equations with the
following choices of freely specifiable data: the spatial conformal
metric is chosen to be flat, the trace of the extrinsic curvature,
its time derivative and the time derivative of the conformal metric
are set to zero.  In
addition to determining the spatial metric and extrinsic curvature,
this form of the initial value equations also determines a lapse and
shift that produce relatively time-independent evolutions.  We use the
Neumann form of the lapse boundary condition described in
Ref.~\cite{Cook2004}.  The binary black holes used for the tests
described here have equal irreducible masses,
$M_{\mathrm{irr}}=1.061536$ (in our code coordinate units), total ADM
energy $M_{\mathrm{ADM}} =2.100609 =1.978834\,M_\mathrm{irr}$, and
total ADM angular momentum $J_{\mathrm{ADM}} =4.3485850
=0.985502\,M^2_\mathrm{ADM}$.  The coordinate separation of the
centers of these holes is set to 20 initially, and the initial spins
of the holes are set to corotating values ({\it i.e.,} no spin in the
corotating frame).  The initial orbital angular velocity of this
binary is $\Omega =0.01418276=0.02979244/M_\mathrm{ADM}$.  This initial 
data set is from a family of publicly available initial 
data~\cite{PublicID}. 

We represent the binary black holes considered here on a computational
domain divided into 44 subdomains: 14 spherical shells, 24 cylindrical
shells, and 6 rectangular blocks.  The various dynamical fields are
expanded in the appropriate spectral basis functions for each
subdomain.  The evolutions described here are performed at two
numerical resolutions using a total (over all subdomains) of
$260,756\approx 64^3$ and $431,566\approx 76^3$ collocation points
respectively.  Constraint preserving and physical boundary conditions
are imposed at the outer boundary of our computational domain
(initially at $r=280$) as described in Ref.~\cite{Lindblom2006}, and
the appropriate characteristic fields are exchanged between subdomains
at internal boundaries.  A more detailed description of the numerical
methods used for these binary black-hole evolutions will be included
in a subsequent paper.  Here we focus our attention on the
dual-coordinate aspects of these evolutions.

As our first binary-black-hole test of the dual-coordinate evolution
method, we evolve the equal-mass circular-orbit binary described above
using a rotating coordinate frame as described in
Sec.~\ref{s:RotatingFrame}.  We set the angular velocity of this frame
to $\Omega =0.02979244/M_\mathrm{ADM}$, the initial orbital angular
velocity of the binary.  We track the position of the center of each
black hole, $[x_c(t), y_c(t), z_c(t)]$, during this evolution by
solving for the apparent horizon of each hole.  We express the
apparent horizon as a spherical-harmonic
expansion~\cite{baumgarte_etal96}: $\vec R_{\mathrm{AH}}= \vec{c}+\hat r
\sum_{\ell,m} R_{lm}(t) Y_{lm}(\theta,\varphi)$, where $\vec c$ is the point about which we expand, $\hat r$ is a
radial unit vector field centered on $\vec c$, and $\theta$ and $\varphi$ label the points
on the apparent-horizon surface.  The coefficients $R_{lm}$ 
are obtained by minimizing the expansion at the Gauss-Legendre collocation 
points with a DFP-minimization~\cite{pfeiffer-etal-2000,Pfeiffer-etal-2002a}. 
The center of the hole is definded as that $\vec c$ for which the $\ell=1$ components of the spherical
harmonic expansion vanish: $R_{1m}=0$.  Initially the black holes are
located at $[x_c(0), y_c(0), z_c(0)]=[\pm 10,0,0]$.  The reflection
symmetry of the initial data ensures that $z_c(t)=0$ throughout the
evolution, but $x_c(t)$ and $y_c(t)$ are free to drift.

\begin{figure} 
\begin{center}
\vspace{0.2cm} \hbox{\hspace{.50cm}
\includegraphics[width=3.0in]{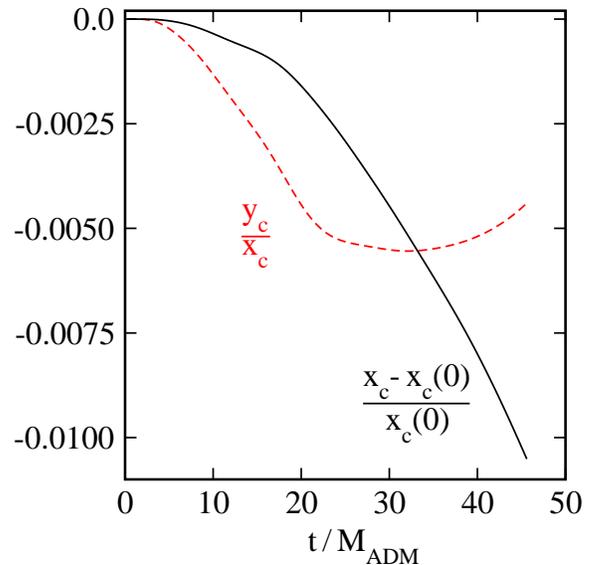} }
\end{center}
\vspace{-0.95cm}
\caption{Evolution of the apparent horizon center, $(x_c,y_c)$, of
one of the black holes relative to a uniformly rotating coordinate frame.
The minimum coordinate distance between the apparent horizon and the
excision boundary becomes less than $0.003$ at $t=45.6\,
M_{\mathrm{ADM}}$, and the evolution is terminated.
\label{f:fig6}}
\vspace{-0.25cm}
\end{figure}
Figure~\ref{f:fig6} illustrates the motion of the center of one of the
black holes, $x_c(t)$ and $y_c(t)$, with respect to the comoving
coordinates.  We see that the center of the hole remains relatively
fixed during the very first part of this evolution, but fairly quickly
it begins to drift away from its initial location with $x_c(t)$
drifting more quickly than $y_c(t)$. It appears that the two black
holes get closer together due to the effects of gravitational
radiation on the orbit.  Our code automatically terminates an
evolution when an apparent horizon gets too close (within 0.003 for
this run) to an inner (excision) boundary of the computational domain.
If the apparent horizon were to cross this boundary then additional
boundary conditions would be needed there and we do not know how to
specify such boundary conditions in a physically meaningful way.

\begin{figure} 
\begin{center}
\vspace{0.2cm}
\hbox{\hspace{0.25cm}
\includegraphics[width=3.0in]{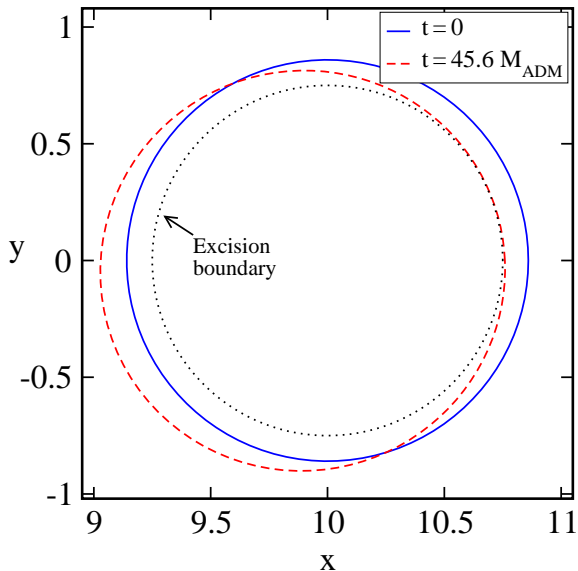}
}
\end{center}
\vspace{-0.95cm}
\caption{Moving coordinate frame representation of an equatorial slice
through the apparent horizon of one of the black holes for the
evolution of Fig.~\ref{f:fig6}.  The evolution is terminated when the
apparent horizon  
approaches the excision boundary (dotted curve) near the right side of this
figure.
\label{f:fig7}}
\vspace{-0.25cm}
\end{figure}
Figure~\ref{f:fig7} shows the equatorial ($x$-$y$ plane) cross
sections of the initial apparent horizon (solid curve), the final
apparent horizon (dashed curve), and the location of the excision
boundary (dotted curve) for one of the black holes
from the evolution shown in Fig.~\ref{f:fig6}.
The
apparent horizon has shifted to the left, confirming our diagnosis that the 
two black holes have gotten closer together.  
Our code does not crash at the end of this test, and the
evolution would have run (a little) longer had a smaller excision
boundary been used.  A smaller excision boundary results in higher
truncation errors for a fixed numerical resolution, so making it
smaller requires significantly more computational resources.  But even
if cost were not an issue, making the excision boundary smaller would only
delay the apparent horizon crossing the excision boundary by a very
brief time.  A better solution is needed for this problem.

\section{Horizon-Tracking Coordinates}
\label{s:HorizonTracking}

The binary black-hole test described in Sec.~\ref{s:BBHtests} shows
that a uniformly-rotating comoving coordinate frame is not adequate
to keep the black holes centered on the excision boundaries of the
computational domain.  The motions of the centers of the holes are too
time-dependent for this simple approach to work effectively for very
long.  If we wish to construct coordinates that track the apparent
horizons for many orbits, then a more flexible coordinate map is
needed.  Since the positions of the black holes are
difficult (impossible) to predict {\it a priori}, some kind of
feedback control system will also be needed to measure these positions
and make the appropriate adjustments to the coordinate maps as the evolution
proceeds.

The motion of the black holes in the test of Sec.~\ref{s:BBHtests} had
a larger $x_c(t)$ component, so we begin by constructing a coordinate map
that more accurately tracks this component of the black hole's
position.  Let us define a control parameter, $Q^{\kern 0.07 em
x}(t)\equiv [x_c(t)-x_c(0)]/x_c(0)$, that gives a dimensionless
measures the $x$-component 
of the position of the
hole.  The idea is to
measure $Q^{\kern 0.07 em x}(t)$ as the binary evolves, and then
use this information to adjust the function $a(t)$ in the
time-dependent coordinate map,
\begin{eqnarray}
\label{e:expandfirstnew}
t&=&\bar t,\\
x&=& a(\bar t\kern 0.15 em)\bigl[ \bar x \cos(\Omega\bar t\kern 0.15 em) 
+ \bar y \sin (\Omega\bar t\kern 0.15 em)\bigr],\\
y&=& a(\bar t\kern 0.15 em)\bigl[ -\bar x \sin(\Omega\bar t\kern 0.15 em) 
+ \bar y \cos( \Omega\bar t\kern 0.15 em)\bigr],\\
\label{e:expandlastnew}
z&=& a(\bar t\kern 0.15 em)\bar z,
\end{eqnarray}
in such a way that $Q^{\kern 0.07 em x}(t)$ remains sufficiently small.
This map applies the uniform rotation used in 
Sec.~\ref{s:BBHtests} combined with a re-scaling of the spatial coordinates 
that can be used to keep the $x$-coordinate separation
of the holes fixed.

We borrow ideas from the literature on mathematical control
theory~\cite{Sontag1998} to design a feedback control system for
the map parameter $a$.  The basic idea is to change $a$ in
such a way that $Q^{\kern 0.07 em x}$ is driven back to its
``equilibrium'' value, $Q^{\kern 0.07 em x}=0$, whenever it drifts
away.  The first step is to determine the response of the control
parameter $\delta Q^{\kern 0.07 em x}$ to a small change in the map
parameter $\delta a$.  In this case the relationship is rather simple:
$\delta Q^{\kern 0.07 em x} = -\delta a/a$.  For maps that
are close to the identity, $a=1$, this relationship is well
approximated by 
$\delta Q^{\kern 0.07 em x} \approx -\delta a$.  We also want to make
sure that any adjustment we make to the coordinate map is sufficiently
smooth that it does not interfere with our ability to solve the
Einstein equations.  In particular we need the coordinate map to be at
least $C^2$ so that the transformed dynamical fields $u^{\bar\alpha}$
are at least continuous [see
Eqs.~(\ref{e:pitrafo})--(\ref{e:newevolsysdef})].  Therefore our
control system will be allowed to adjust only $d^{\kern 0.06em
3}a/dt^3$ freely, and $a$ will be obtained by time integration
to ensure sufficient smoothness.
Following the usual procedure in control
theory, we choose $d^{\kern 0.06em 3}a/dt^3$ in a way that is
determined by the measured values of $Q^{\kern 0.07 em x}$ and its
derivatives:
\begin{eqnarray}
\frac{d^{\kern 0.06em 3}a}{dt^3} = 
\alpha Q^{\kern 0.07 em x} + \beta \frac{dQ^{\kern 0.07 em x}}{dt} 
+\gamma \frac{d^{\kern 0.06em 2}Q^{\kern 0.07 em x}}{dt^2} ,
\label{e:control} 
\end{eqnarray}
where $\alpha$, $\beta$ and $\gamma$ are parameters that we are free to pick.
We choose these parameters in such a way that the ``closed-loop''
equation,
\begin{eqnarray}
\frac{d^{\kern 0.06em 3}a}{dt^3} = -\frac{d^{\kern 0.06em 3}
Q^{\kern 0.07 em x}}{dt^3} =
\alpha Q^{\kern 0.07 em x} + \beta \frac{dQ^{\kern 0.07 em x}}{dt} 
+\gamma \frac{d^{\kern 0.06em 2}Q^{\kern 0.07 em x}}{dt^2} ,
\label{e:closedloop} 
\end{eqnarray}
has solutions that all decay exponentially toward the desired equilibrium
value $Q^{\kern 0.07 em x}\rightarrow 0$.  We use the values
$\alpha=\lambda^3$, $\beta=3\lambda^2$ and $\gamma=3\lambda$,
that result in the following closed-loop equation:
\begin{eqnarray}
\frac{d^{\kern 0.06em 3}Q^{\kern 0.07 em x}}{dt^3} =
-\lambda^3 Q^{\kern 0.07 em x} - 3 \lambda^2 \frac{dQ^{\kern 0.07 em x}}{dt} 
- 3\lambda \frac{d^{\kern 0.06em 2}Q^{\kern 0.07 em x}}{dt^2} .
\label{e:controlfinal} 
\end{eqnarray}
The most general solution to this equation is
\begin{eqnarray}
Q^{\kern 0.07 em x}(t)=(At^2+Bt+C)e^{-\lambda t},
\end{eqnarray}
for arbitrary constants $A$, $B$, and $C$.  All of these solutions
decay exponentially toward the desired equilibrium solution at a rate
determined by the parameter $\lambda>0$.

We use these ideas now to construct a feedback control system for
$a$.  We pick a set of control times $t_i$ at which the
expression for $a$ will be adjusted.  The control interval
$\Delta t=t_{i+1}-t_{i}$ is chosen to be shorter than the timescale on
which $Q^{\kern 0.07 em x}$ drifts away from its equilibrium value.
We choose the map parameter $a(t)$ in the time interval $t_i\leq
t < t_{i+1}$ to be:
\begin{eqnarray}
a(t)&=& a_i + (t-t_i)\frac{da_i}{dt}
+\frac{(t-t_i)^2}{2}\frac{d^{\kern 0.06em 2}a_i}{dt^2}\nonumber\\
&&+\frac{(t-t_i)^3}{2}\left(
\lambda\frac{d^{\kern 0.06 em 2}Q^{\kern 0.07 em x}_i}{dt^2}
+\lambda^2\frac{dQ^{\kern 0.07 em x}_i}{dt}
+\lambda^3 \frac{Q^{\kern 0.07 em x}_i}{3}\right),\quad
\label{e:controla}
\end{eqnarray}
where the constants $a_i$, $da_i/dt$ and $d^{\kern 0.06em 2}a_i/dt^2$
are the values taken from the map in the previous interval
$t_{i-1}\leq t < t_i$ evaluated at $t=t_i$, and the constants
$Q^{\kern 0.07 em x}_i$, $dQ^{\kern 0.07 em x}_i/dt$ and $d^{\kern
0.06 em 2}Q^{\kern 0.07 em x}_i/dt^2$ are the values measured at
$t=t_i$.  This choice of $a(t)$ guarantees that our comoving
coordinate map is $C^2$ in time, $C^\infty$ in space, and it enforces
the closed-loop equation exactly at the control times $t=t_i$.  We
begin our evolutions at $t_0=0$ by setting the initial conditions
$a_0=1$ and $da_0/dt = d^{\kern 0.06em 2}a_0/dt^2 = Q^{\kern 0.07 em
x}_0 = dQ^{\kern 0.07 em x}_0/dt = d^{\kern 0.06em 2}Q^{\kern 0.07 em
x}_0/dt^2 = 0$.  To ensure that the discontinuities in $d^{\kern
0.06em 3}a/dt^3$ do not affect the convergence of our code, we require
that the $t_i$ occur at fixed times (independent of the numerical
timestep), and we choose the timesteps so that the $t_i$ always
coincide with the beginning of a full numerical timestep.

\begin{figure} 
\begin{center}
\vspace{0.2cm}
\hbox{\hspace{.50cm}
\includegraphics[width=3.0in]{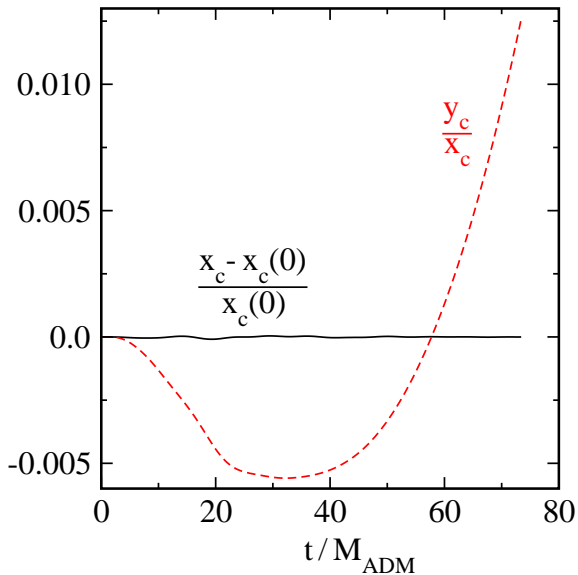}
}
\end{center}
\vspace{-0.95cm}
\caption{Evolution of the apparent horizon center of one of the holes
for a binary
black-hole evolution using a feedback control system that adjusts the moving
coordinate frame to control the parameter $[x_c-x_c(0)]/x_c(0)$.  Compare to Fig.~\ref{f:fig6}.
\label{f:fig8}}
\vspace{-0.25cm}
\end{figure}
\begin{figure} 
\begin{center}
\vspace{0.2cm}
\hbox{\hspace{0.25cm}
\includegraphics[width=3.0in]{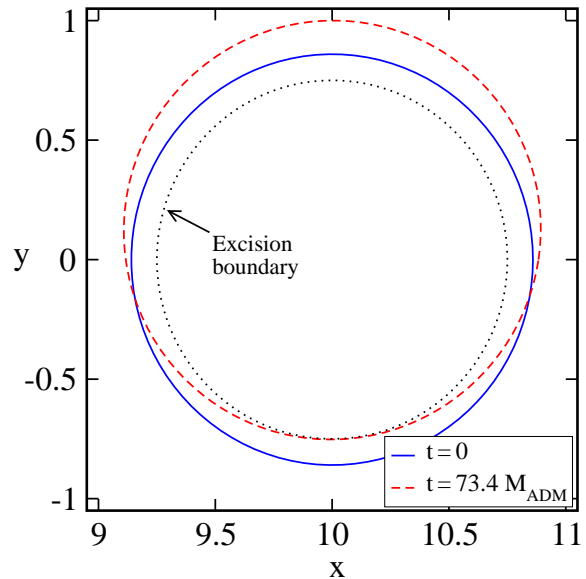}
}
\end{center}
\vspace{-0.95cm}
\caption{Moving coordinate frame representation of an equatorial slice
through the apparent horizon of one of the black holes for the evolution of
Fig.~\ref{f:fig8}.  Evolution is terminated when the apparent horizon
approaches the excision boundary near the bottom of this figure.
\label{f:fig9}}
\vspace{-0.25cm}
\end{figure}

We test this feedback control system by evolving the same binary
black-hole initial data discussed in Sec.~\ref{s:BBHtests}.  We set
the value of the control damping parameter $\lambda \approx
0.3/M_\mathrm{ADM}$, the control interval $\Delta t\approx 1\,
M_\mathrm{ADM}$, and the timescale used to evaluate average values of 
the derivatives of
$Q^{\kern 0.07 em x}$ in Eq.~(\ref{e:controla}) to $\sim
0.5\,M_\mathrm{ADM}$.  Figure~\ref{f:fig8} shows the motion of the
center of one black hole during this test.  We see that the control
system effectively keeps the $x$-coordinate of the center of the hole
$Q^{\kern 0.07 em x}=[x_c-x_c(0)]/x_c(0)$ within acceptable bounds.  
But this test
ends at about $t= 73.4\, M_\mathrm{ADM}$ when the other (uncontrolled)
component of the black hole's position $y_c/x_c$ grows too
large.  The new control system significantly extends this binary
evolution, but the binary has still only completed about 0.34 orbits.
Figure~\ref{f:fig9} shows that the apparent horizon has drifted upward
at the time this test ends.  As the binary inspirals
towards merger the angular velocity of the orbit increases, so
the fixed angular velocity of the co-moving coordinates is no longer
able to track the positions of the holes with sufficient accuracy.

Fortunately, the dual-coordinate evolution method is extremely flexible
and it is easy to construct a comoving coordinate map capable of
tracking both the orbital rotation and the radial motion of the black
holes.  For example, the map
\begin{eqnarray}
\label{e:expandandrotfirstnew}
t&=&\bar t,\\
x&=& a(\bar t\kern 0.15 em)\bigl[ \bar x \cos\varphi(\bar t\kern 0.15 em) 
+ \bar y \sin \varphi(\bar t\kern 0.15 em)\bigr],\\
y&=& a(\bar t\kern 0.15 em)\bigl[ -\bar x \sin\varphi(\bar t\kern 0.15 em) 
+ \bar y \cos \varphi(\bar t\kern 0.15 em)\bigr],\\
\label{e:expandandrotlastnew}
z&=& a(\bar t\kern 0.15 em)\bar z
\end{eqnarray}
includes a time-dependent rotation angle $\varphi(t)$ that can be
used to adjust the angular velocities of the holes, in
addition to the time-dependent conformal factor $a(t)$.  An
additional feedback control system is then needed to adjust the
map parameter $\varphi$ through measurements of the control parameter
$Q^{\kern 0.07 em y}\equiv y_c/x_c$.  The basic control
equation for this parameter is $\delta Q^{\kern 0.07 em y} = -\delta
\varphi$.  So we construct a feedback control system 
for $\varphi(t)$ that is completely analogous to Eq.~(\ref{e:controla}).
In the time interval $t_i\leq t < t_i$ we set
\begin{eqnarray}
\varphi(t)&=& \varphi_i + (t-t_i)\frac{d\varphi_i}{dt}
+\frac{(t-t_i)^2}{2}\frac{d^{\kern 0.06em 2}\varphi_i}{dt^2}\nonumber\\
&&+\frac{(t-t_i)^3}{2}\left(
\lambda\frac{d^{\kern 0.06 em 2}Q^{\kern 0.07 em y}_i}{dt^2}
+\lambda^2\frac{dQ^{\kern 0.07 em y}_i}{dt}
+\lambda^3 \frac{Q^{\kern 0.07 em y}_i}{3}\right),\quad
\label{e:controlphi}
\end{eqnarray}
where the constants $\varphi_i$, $d\varphi_i/dt$ and 
$d^{\kern 0.06em 2}\varphi_i/dt^2$
are the values taken from the map in the previous interval
$t_{i-1}\leq t < t_i$ evaluated at $t=t_i$, and the constants
$Q^{\kern 0.07 em y}_i$, $dQ^{\kern 0.07 em y}_i/dt$ and $d^{\kern
0.06 em 2}Q^{\kern 0.07 em y}_i/dt^2$ are the values measured at
$t=t_i$.  This choice of $\varphi(t)$ guarantees that our comoving
coordinate map is $C^2$, and it enforces the closed-loop equation
periodically at the control times $t=t_i$.  We begin our evolutions at
$t_0=0$ by setting the initial conditions $d\varphi_0/dt=\Omega$ and 
$\varphi_0 =
d^{\kern 0.06em 2}\varphi_0/dt^2 = Q^{\kern 0.07 em y}_0 = dQ^{\kern 0.07 em
y}_0/dt = d^{\kern 0.06em 2}Q^{\kern 0.07 em y}_0/dt^2 = 0$.

\begin{figure} 
\begin{center}
\vspace{0.2cm}
\hbox{\hspace{.50cm}
\includegraphics[width=3.0in]{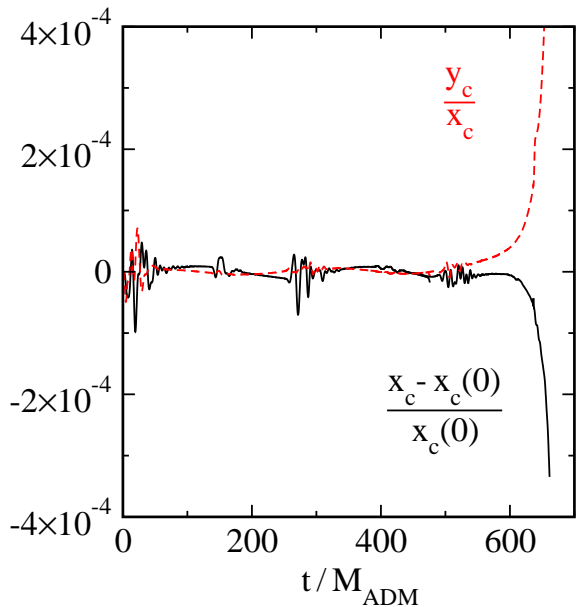}
}
\end{center}
\vspace{-0.95cm}
\caption{Evolution of the apparent horizon center of one black hole
for a binary
black-hole evolution using a feedback system that adjusts the moving
coordinate frame to control both the parameters $y_c/x_c$
and $[x_c-x_c(0)]/x_c(0)$.  Compare to Figs.~\ref{f:fig6} and~\ref{f:fig8}.
\label{f:fig10}}
\vspace{-0.25cm}
\end{figure}
\begin{figure} 
\begin{center}
\vspace{0.2cm}
\hbox{\hspace{.50cm}
\includegraphics[width=3.0in]{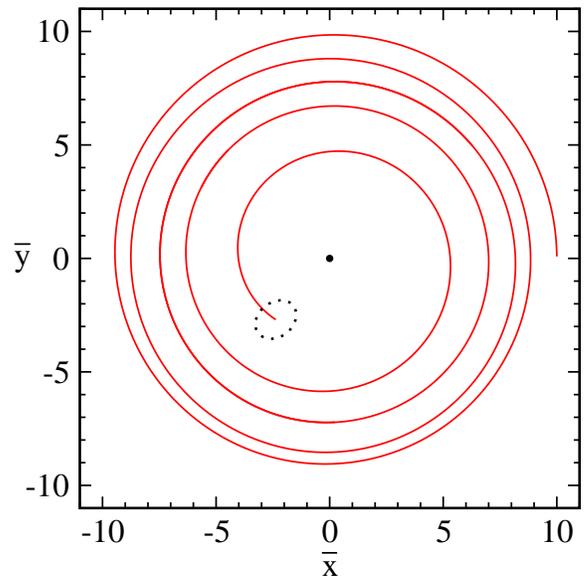}
}
\end{center}
\vspace{-0.95cm}
\caption{Motion of the inertial frame coordinates of the center of
the apparent horizon $(\bar x_c, \bar y_c)$ (solid curve) of one
of the black holes from the evolution of Fig.~\ref{f:fig10}.  The dotted
curve shows the equatorial cross section of the apparent horizon at
$t=661.4\,M_\mathrm{ADM}$.  
The binary system has completed about 4.6 orbits.
\label{f:fig11}}
\vspace{-0.25cm}
\end{figure}
\begin{figure} 
\begin{center}
\vspace{0.2cm}
\hbox{\hspace{.50cm}
\includegraphics[width=3.0in]{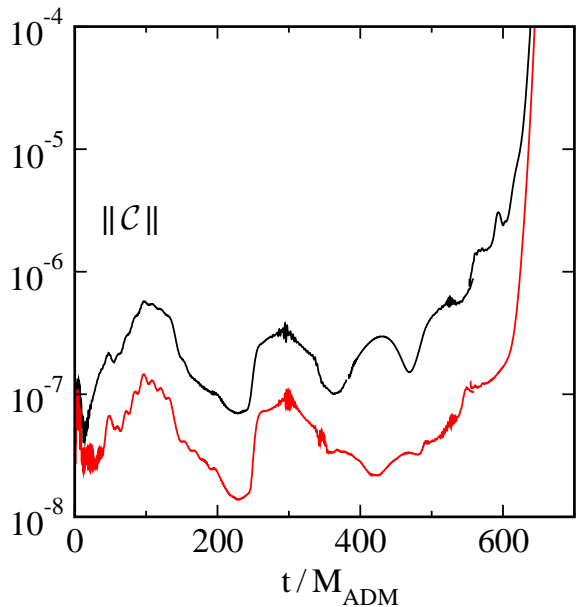}
}
\end{center}
\vspace{-0.95cm}
\caption{Constraint violations $||\,{\cal C}\,||$ for the binary
black-hole evolution of Fig.~\ref{f:fig10}.  The two curves represent
two numerical resolutions, the upper curve has about $64^3$ collocation
points, while the lower curve has about $76^3$ collocation points. 
\label{f:fig12}}
\vspace{-0.25cm}
\end{figure}

We test this enhanced feedback control system, which controls both
$Q^{\kern 0.07 em x}$ and $Q^{\kern 0.07 em y}$, by evolving the same
binary black-hole initial data used for the evolutions in
Figs.~\ref{f:fig6} and \ref{f:fig8}.  We use the same control system
parameters for the $Q^{\kern 0.07 em x}$ subsystem as those discussed
above.    For the $Q^{\kern 0.07 em y}$ subsystem we use
the value of the control damping parameter $\lambda \approx
0.5/M_\mathrm{ADM}$, the control interval $\Delta
t\approx 0.5\, M_\mathrm{ADM}$, and the timescale used to evaluate the
average values of the 
derivatives of $Q^{\kern 0.07 em y}$ in Eq.~(\ref{e:controlphi}) to
$\sim 0.2\,M_\mathrm{ADM}$.  
Figure~\ref{f:fig10} shows the control parameters $Q^{\kern 0.07 em
x}$ and $Q^{\kern 0.07 em y}$ for this evolution.  We see that both
control systems work extremely well, and the binary now evolves until
$t=661.4\,M_\mathrm{ADM}$ before the evolution stops.
Figure~\ref{f:fig11} shows the evolution of the {\it inertial-frame}
coordinates $[\bar x_c(\bar t\kern 0.15em), \bar y_c(\bar t\kern
0.15em)]$ of the center of one of the black holes, from which we see 
that this evolution has completed about 4.6 orbits.  
Figure~\ref{f:fig12} shows the
normalized constraint violations for this evolution, at two different
numerical resolutions.  The upper curve uses a total of
$260,756\approx 64^3$ collocation points and the lower curve uses
$431,566\approx 76^3$ collocation points.  The constraints are
normalized here by dividing by the {\it initial} value of the norm of
the derivatives of the dynamical fields.  Figure~\ref{f:fig12}
suggests that our numerical evolutions are convergent until just before
the code terminates.

\begin{figure} 
\begin{center}
\vspace{0.2cm}
\hbox{\hspace{0.25cm}
\includegraphics[width=3.0in]{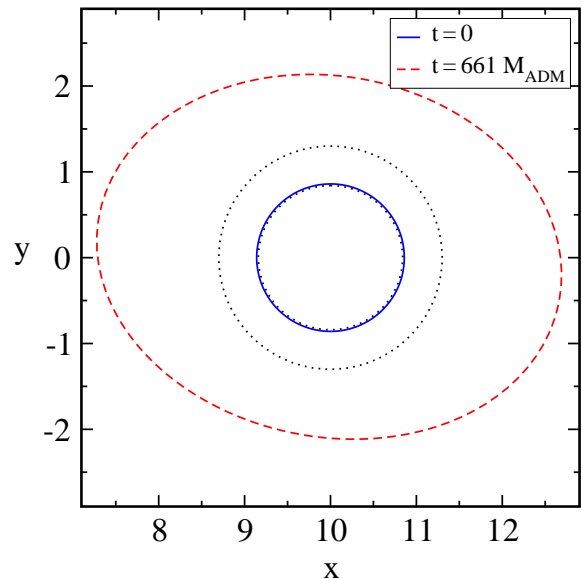}
}
\end{center}
\vspace{-0.95cm}
\caption{Equatorial cross sections, in comoving coordinates,
of the apparent horizon of one hole for the evolution of Fig.~\ref{f:fig10}.
The horizon is shown both at the beginning and at the end of the evolution. 
The smaller dotted circle (just inside the initial horizon) shows the
excision boundary used for $t \leq 554.1\,M_\mathrm{ADM}$. The outer
dotted circle shows the excision boundary used for $t>
554.1\,M_\mathrm{ADM}$.
\label{f:fig13}}
\vspace{-0.25cm}
\end{figure}

Finally, Fig.~\ref{f:fig13} shows the shape of the apparent horizon of
one black hole (expressed in comoving coordinates) at $t=0$ (solid
curve) and at the end of the evolution $t=661.4\,M_\mathrm{ADM}$
(dashed curve).  The conformal factor $a(t)$ used to keep the black
holes at a fixed coordinate separation also causes the apparent
horizons to expand in comoving coordinates.  As the binary evolves, the
excision boundary (which has a fixed coordinate radius in the moving frame) 
moves deeper
and deeper into the interior of the black hole.  At fixed numerical
resolution this leads to rapidly increasing truncation errors that in
turn generate constraint violations. The dominant source of constraint
violations seen at the end of the evolutions in Fig.~\ref{f:fig12}
comes from the region within the apparent horizons of the two black
holes.  This problem was partially corrected by moving the location of
the excision boundaries at the time $t=554.1\,M_\mathrm{ADM}$: for each
hole, a portion
of the unphysical interior grid was removed (by interpolating the
dynamical fields onto a smaller grid), thus moving the excision
boundary to the location of the outer dotted curve seen in
Fig.~\ref{f:fig13}.  This evolution stops in part because truncation
errors grow rapidly as the excision boundaries move deeper inside the
holes.  Another reason this evolution stops can also be seen in
Fig.~\ref{f:fig13}: the shape of each apparent horizon has become very
distorted by tidal interaction with the companion hole.  This
distortion requires a significant increase in angular resolution to
represent it accurately; so at fixed numerical resolution, truncation
errors quickly grow and these cause rapidly growing constraint
violations seen near the end of the evolutions in Fig.~\ref{f:fig12}.

\section{Discussion}
\label{s:Discussion}

We have shown in Sec.~\ref{s:RotatingFrame} that the dual-coordinate
frame evolution method introduced in Sec.~\ref{s:Moving Frames} is
capable of solving the rotating-frame instability problem.  A more
sophisticated application of this method has also allowed us to
construct comoving coordinates in Sec.~\ref{s:HorizonTracking} that
track the apparent horizons of a binary black hole system through a
feedback control system.  These horizon-tracking coordinates allowed
us to evolve a binary black-hole system stably and accurately for
about 4.6 orbits.  Our expectation is that the great flexibility of
the dual-coordinate method will make it useful in many other
applications.  We imagine, for example, that coordinate maps
could be constructed that keep the shape of the apparent horizons
spherical, and their locations close to the excision boundary.  Such
maps should solve the problems that prevent the binary black-hole
evolution shown here from proceeding all the way to merger.  We also
expect that many applications are possible that we are not able to
anticipate at this point.

\section*{Appendix}
\label{s:Appendix}

In this appendix, we describe some of our efforts to understand and
cure the instability that occurs when solving the generalized harmonic
system using a single rotating coordinate frame.  We found the
instability, illustrated in Fig.~\ref{f:fig1}, to occur even for
evolutions of flat spacetime in a rotating frame.  Since Minkowski
space is simpler than Schwarzschild, much of our analysis was focused
on understanding the flat-space case.

At the analytical level, it is easy to see that in Minkowski space 
stability in non-rotating inertial coordinates is equivalent to
stability in a single rotating coordinate frame.  The
basic argument is as follows.  Consider two coordinate systems
$x^{\bar a}$ and $x^a$ related by a uniform rotation,
Eqs.~(\ref{e:rotfirst})--(\ref{e:rotlast}).  The evolution equations for
linear perturbations of the metric $\delta\psi_{ab}$ about a flat
Minkowski background may be written in an abstract way as $E_{ab}
[\delta\psi] = 0$.  We assume that
both the fields and these evolution equations transform as
tensors, {\it i.e.,} $\delta\psi_{ab} = A_a{}^{\bar a}A_b{}^{\bar b}
\delta\psi_{\bar a\bar b}$ and $E_{ab} = A_a{}^{\bar a}A_b{}^{\bar b}
E_{\bar a\bar b}$, where $A_a{}^{\bar a} \equiv \partial x^{\bar a} /
\partial x^a$.  This implies that if $\delta\psi_{\bar a\bar b}$
is a solution in the inertial frame then $\delta\psi_{ab} =
A_a{}^{\bar a}A_b{}^{\bar b} \delta\psi_{\bar a\bar b}$ is a solution
in the rotating frame.  Suppose that $\delta \psi_{\bar a\bar b} \sim
e^{s \bar t} e^{i m \bar \varphi}$ is an eigenmode in the inertial
frame. Then $\delta\psi_{ab} \sim e^{s' t} e^{i m \varphi}$ is a
solution in the rotating frame with $s' = s - i m \Omega$, so that
$\Re(s') = \Re(s)$.  Thus, linear stability in the two frames is
equivalent.  We have done a complete linear stability analysis
in the inertial frame, and the solutions are (of course) stable.  
 Nevertheless, our generalized harmonic evolution code is
stable for nearly flat evolutions in non-rotating coordinates, but is
extremely unstable for solutions in a rotating frame.

To understand this apparent contradiction, let us now consider the second-order
form of the vacuum
generalized harmonic evolution equations in more detail,
\begin{equation}
  \label{e:gheqns}
  E_{ab}[\psi] = R_{ab}[\psi] - \nabla_{(a} \mathcal{C}_{b)}[\psi] = 0,
\end{equation}
where $\psi_{ab}$ is the four-metric, $\nabla_a$ its covariant derivative,
$R_{ab}$ the Ricci tensor, and $\mathcal{C}_a$ is the harmonic constraint
defined in Eq.~(\ref{e:ghconstr}).
Equation~(\ref{e:gheqns}) transforms as a
tensor if and only if $\mathcal{C}_a$ transforms as a vector. For this
to be true, we have to demand a rather special transformation law for
the gauge source function $H_a$: from the known transformation of the
Christoffel symbols and Eq.~(\ref{e:ghconstr}) we deduce
\begin{equation}
  \label{e:htrafo}
  H_a = \tilde H_{a} - \psi_{ad} \psi^{bc} \hat \Gamma^d{}_{bc}, 
\end{equation}
where $\tilde H_a \equiv A_a{}^{\bar a} H_{\bar a}$ and  
$\hat \Gamma^a{}_{bc} \equiv (A^{-1})^a{}_{\bar a} \partial_c
A_b{}^{\bar a}$.  Naively, in order to evolve Schwarzschild spacetime
in a rotating frame, say, one would just set $H_a$ to be the
$-\Gamma_a$ of the exact solution evaluated in that frame and freeze it
in time. In contrast, Eq.~(\ref{e:htrafo}) tells us that $H_a$ should
depend on the evolved metric $\psi_{ab}$.  We find that
freezing the quantity $\tilde H_a$ in the rotating frame
and then obtaining $H_a$ from Eq.~\eqref{e:htrafo} using the evolved
metric $\psi_{ab}$ makes our rotating-frame Minkowski
and Schwarzschild evolutions much more (but not completely)
stable.

Our numerical code solves a first-order representation of the
generalized harmonic evolution equations.  To understand the remaining
problems we must look at the first-order reduction in more detail.
The first-order formulation introduces the new dynamical fields
\begin{equation}
  \label{e:1ovars1}
  \Pi_{ab} \equiv - t^c \partial_c \psi_{ab}, \quad
  \Phi_{kab} \equiv \partial_k \psi_{ab}.
\end{equation}
Unlike the metric $\psi_{ab}$, these new fields do not transform as
tensors, as can be appreciated from
Eqs.~(\ref{e:pitrafo})--(\ref{e:phitrafo}).  Thus the simple analytical
argument given above for the stability of rotating flat space does not
apply.  We can modify the standard first-order form of the equations
by changing the definitions of these new fields slightly. We replace
Eq.~(\ref{e:1ovars1}) with
\begin{equation}
  \label{e:1ovars2}
  {\tilde \Pi}_{ab} \equiv - t^c {\hat \nabla}_c \psi_{ab}, \quad
  {\tilde \Phi}_{kab} \equiv {\hat \nabla}_k \psi_{ab},  
\end{equation}
where $\hat \nabla_a$ denotes the covariant derivative associated with
the connection $\hat \Gamma^a{}_{bc}$ introduced in
Eq.~(\ref{e:htrafo}).  By definition, $\hat \nabla_a$ reduces to
$\partial_a$ in the inertial frame, but not in the rotating frame.
The modified variables in Eq.~(\ref{e:1ovars2}) by definition transform as
tensors under rotations of the coordinates.  To make this new
first-order formulation truly covariant we must also ensure that the
evolution equations (and boundary conditions) transform as tensors. To
do this, we start with the equations written in the inertial frame
(see {\it e.g.,} Eqs.~[35]--[37] of Ref.~\cite{Lindblom2006}) and
formally replace all partial derivatives $\partial_a$ with the
covariant derivatives $\hat \nabla_a$.  The rotating frame
solutions of these modified evolution equations should then have the
same stability properties as the solutions obtained in the inertial
frame.  We have used this system to evolve both Minkowski and
Schwarzschild spacetimes, and found the numerical solutions to be 
greatly improved over the unmodified system.  But the results are still
not quite stable.

The next problem that we discovered turned out to be numerical: we
needed to modify the standard tensor-spherical-harmonic filtering
algorithm described for example in Sec.~\ref{s:RotatingFrame}.  The
transformation that relates the rotating frame components of tensors
with non-rotating frame components mixes together the time components
with some spatial components ({\it e.g.,} $\psi_{tt}$ depends on
$\psi_{\bar t\bar t}$, $\psi_{\bar t\bar \varphi}$ and
$\psi_{\bar\varphi\bar\varphi}$).  This causes scalar, vector, and
second-rank tensor spherical harmonics to be mixed together in a
non-linear way.  To disentangle these parts, it is necessary first to
transform the time components of these fields to the inertial frame,
filter them there, and then transform them back to the rotating
frame. In this way, we remove the mixing of the different tensor
spherical harmonics that occurs in a rotating frame.  Using this
filtering method (which turns out to be equivalent to the filtering
used for our dual-coordinate evolutions described in
Sec.~\ref{s:RotatingFrame}) produces Minkowski-space evolutions that
appear to be completely stable, while evolutions of Schwarzschild
though improved are still not quite stable.

The combination of the modifications described above significantly
improves the stability of our evolutions of flat space and single
black-hole spacetimes in rotating coordinates.  However, these methods
did not produce robustly stable evolutions over the entire range of
domain sizes and rotating frame angular velocities needed for binary
black hole simulations.  We stopped pursuing this approach after the
dual-frame method was found to be so stable, and so flexible that it
allowed us to solve our moving frame excision problem as well.

\acknowledgments This work was supported in part by a grant from the
Sherman Fairchild Foundation to Caltech and Cornell, by NSF grants
PHY-0099568, PHY-0244906, PHY-0601459 and NASA grants NAG5-12834,
NNG05GG52G at Caltech, and by NSF grants PHY-0312072, PHY-0354631, and
NASA grant NNG05GG51G at Cornell.

\bibliography{References}

\end{document}